\documentclass[10pt]{article}
\usepackage{graphicx}
\usepackage{amsmath}
\usepackage{amssymb}
\usepackage{caption2}
\begin{document}
\bibliographystyle{prsty}
\begin{center}
{\large {\bf \sc{  The lowest hidden charmed  tetraquark state
from  QCD sum rules }}} \\[2mm]
Zhi-Gang  Wang \footnote{E-mail: zgwang@aliyun.com.  }     \\
 Department of Physics, North China Electric Power University, Baoding 071003, P. R. China
\end{center}

\begin{abstract}
In this article, we study the   $S\bar{S}$ type scalar tetraquark state $cq\bar{c}\bar{q}$ in details with the QCD sum rules by calculating the contributions of the vacuum condensates up to dimension-10  in the operator product expansion, and obtain the value $M_{Z_c}=\left(3.82^{+0.08}_{-0.08}\right)\,\rm{GeV}$, which is the lowest mass for the hidden charmed tetraquark states from the QCD sum rules.  Furthermore, we calculate  the hadronic coupling constants $G_{Z_c\eta_c\pi}$ and $G_{Z_cDD}$ with the three-point QCD sum rules, then study the strong decays  $ Z_c\to \eta_c\pi\, ,\, DD$, and observe that the total width $\Gamma_{Z_c}\approx 21\,\rm{MeV}$.  The present
  predictions can be confronted with the experimental data in the futures at the BESIII, LHCb and Belle-II.
\end{abstract}



\section{Introduction}

The scattering amplitude for one-gluon exchange in an $SU(N_c)$
gauge theory is proportional to
\begin{eqnarray}
t^a_{ki}t^a_{lj}&=&-\frac{N_c+1}{4N_c}(\delta_{jk}\delta_{il}-\delta_{ik}\delta_{jl})
 +\frac{N_c-1}{4N_c}(\delta_{jk}\delta_{il}+\delta_{ik}\delta_{jl})\, ,
\end{eqnarray}
where the $t^a$ is the generator of the gauge group, and the $i,j$
and $k,l$ are the color indexes of the two quarks in the incoming
and outgoing channels respectively.   For $N_c=3$, the negative sign
in front of the antisymmetric  antitriplet indicates the interaction
is attractive and favors the formation of the diquark states in the color
 antitriplet, while the positive sign in front of the symmetric
sextet indicates
 the interaction  is repulsive and disfavors the formation of the diquark states in the color
 sextet \cite{Huang-2005}.

 The antitriplet diquark states have  five Dirac tensor structures, scalar $C\gamma_5$,
pseudoscalar $C$, vector $C\gamma_\mu \gamma_5$, axial vector
$C\gamma_\mu $  and  tensor $C\sigma_{\mu\nu}$. The structures
$C\gamma_\mu $ and $C\sigma_{\mu\nu}$ are symmetric, the structures
$C\gamma_5$, $C$ and $C\gamma_\mu \gamma_5$ are antisymmetric. The
attractive interactions of one-gluon exchange  favor  formation of
the diquarks in  color antitriplet $\overline{3}_{ c}$, flavor
antitriplet $\overline{3}_{ f}$ and spin singlet $1_s$ (or flavor
sextet  $6_{ f}$ and spin triplet $3_s$) \cite{One-gluon-1,One-gluon-2}, so the favored configurations are the scalar and axial-vector diquark states.
The scalar ($S$) and axial-vector ($A$) heavy-light diquark states have almost  degenerate masses from the QCD sum rules \cite{WangDiquark-1,WangDiquark-2}.
In Refs.\cite{Wang-Scalar-1,Wang-Scalar-2}, we take the $C\gamma_5-C\gamma_5$, $C\gamma_\mu-C\gamma^\mu$, $C\gamma_\mu\gamma_5-C\gamma^\mu\gamma_5$ type interpolating currents to study the masses of the scalar tetraquark states in a systematic way using the QCD sum rules, and observe that  the $S\bar{S}$ and $A\bar{A}$ type scalar tetraquark states have almost degenerate masses, about $4.36\,\rm{GeV}$, which is much larger than that from the phenomenological models \cite{EFG-2008,Maiani-2004,Maiani-2014}.

In Ref.\cite{EFG-2008},   Ebert,  Faustov and   Galkin calculate the masses of the excited heavy tetraquarks with hidden charm
within the relativistic diquark-antidiquark picture based on the quasipotential approach, and obtain the values $M_{J=0}=3.852\,\rm{GeV}$ and $3.812\,\rm{GeV}$ for the $A\bar{A}$ and $S\bar{S}$ type scalar tetraquark states $cq\bar{c}\bar{q}$, respectively. While L. Maiani et al obtain the values $M_{J=0}=3.832\,\rm{GeV}$ and $3.723\,\rm{GeV}$  for the $A\bar{A}$ and $S\bar{S}$ type scalar tetraquark states $cq\bar{c}\bar{q}$ respectively in the type-I diquark-antidiquark model \cite{Maiani-2004}, and
$M_{J=0}=4.000\,\rm{GeV}$ and $3.770\,\rm{GeV}$ for the $A\bar{A}$ and $S\bar{S}$ type scalar tetraquark states $cq\bar{c}\bar{q}$ respectively in the type-II diquark-antidiquark model \cite{Maiani-2014}. In those  model-dependent studies, the  masses of the $A\bar{A}$-type scalar tetraquark states are larger  than that of the $S\bar{S}$ type scalar tetraquark states.

In Refs.\cite{WangHuangTao-1,WangHuangTao-2,WangHuangTao-3,Wang-Cu-Cu,WangHuang-molecule-1,WangHuang-molecule-2},   we    explore the energy scale dependence of the hidden charmed (bottom) tetraquark states and molecular states  in details for the first time, and suggest a  formula
\begin{eqnarray}
\mu&=&\sqrt{M^2_{X/Y/Z}-(2{\mathbb{M}}_Q)^2} \, ,
 \end{eqnarray}
 with the effective heavy $Q$-quark mass ${\mathbb{M}}_Q$ to determine the energy scales of the  QCD spectral densities in the QCD sum rules, which works well.
According to the formula, the energy scale $\mu=1\,\rm{GeV}$ taken in Refs.\cite{Wang-Scalar-1,Wang-Scalar-2} is too low to result in  robust predictions.

In Ref.\cite{Wang-Cu-Cu}, we choose the $C\gamma_\mu-C\gamma_\nu$  type  interpolating currents to study the $A\bar{A}$-type  scalar, axial-vector and tensor tetraquark states in details with the QCD sum rules.  The predicted  masses of the axial-vector and tensor tetraquark states   favor assigning  the $Z_c(4020)$ and $Z_c(4025)$   as the $J^{PC}=1^{+-}$ or $2^{++}$   diquark-antidiquark type tetraquark states.  While there are no experimental candidates to match  the predicted  mass of the scalar tetraquark state $M_{J=0}=\left(3.85^{+0.15}_{-0.09}\right)\,\rm{GeV}$. The value is consistent with the prediction $M_{J=0}=3.852\,\rm{GeV}$ based on the quasipotential approach \cite{EFG-2008}, while the upper bound reaches the prediction $M_{J=0}=4.000\,\rm{GeV}$ based on  the type-II diquark-antidiquark model \cite{Maiani-2014}. According to Refs.\cite{EFG-2008,Maiani-2004,Maiani-2014}, the $S\bar{S}$-type scalar tetraquark states have smaller masses than that of the corresponding $A\bar{A}$-type scalar tetraquark states. It is interesting to see whether or not such conclusion survives when confronted  with the QCD sum rules. In Refs.\cite{WangHuangTao-1,WangHuangTao-2,WangHuangTao-3}, we observe that the masses of the $S\bar{A}$ or $A\bar{S}$ type axial-vector tetraquark states are larger than that of the $A\bar{A}$ type scalar tetraquark states. So the $S\bar{S}$ scalar tetraquark state maybe  the lowest tetraquark state.

 In this article, we study the  scalar $S\bar{S}$-type  hidden charmed tetraquark state (thereafter we will denote it as $Z_c$) by calculating the contributions of the vacuum condensates up to dimension-10, and try to obtain the lowest mass based on the QCD sum rules. Furthermore, we calculate  the hadronic coupling constants $G_{Z_c\eta_c\pi}$ and $G_{Z_cDD}$ with the three-point QCD sum rules, then study the strong decays  $ Z_c\to \eta_c\pi\, ,\, DD$.

The article is arranged as follows:  we derive the QCD sum rules for the mass and pole residue of  the scalar  tetraquark state $Z_c$ and for  the hadronic coupling constants $G_{Z_c\eta_c\pi}$ and $G_{Z_cDD}$  in section 2; in section 3, we present the numerical results and discussions; section 4 is reserved for our conclusion.

\section{QCD sum rules for  the  scalar tetraquark state }
In the following, we write down  the two-point correlation function $\Pi(p)$ in the QCD sum rules,
\begin{eqnarray}
\Pi(p)&=&i\int d^4x e^{ip \cdot x} \langle0|T\left\{J(x)J^{\dagger}(0)\right\}|0\rangle \, ,
\end{eqnarray}
\begin{eqnarray}
   J(x)&=&\epsilon^{ijk}\epsilon^{imn}u^j(x)C\gamma_5 c^k(x) \bar{d}^m(x)\gamma_5 C \bar{c}^n(x) \, ,
\end{eqnarray}
where the $i$, $j$, $k$, $m$, $n$ are color indexes, the $C$ is the charge conjugation matrix.

At the hadronic side, we can insert  a complete set of intermediate hadronic states with
the same quantum numbers as the current operator  $J(x)$ into the
correlation function  $\Pi(p)$ to obtain the hadronic representation
\cite{SVZ79-1,SVZ79-2,Reinders85}. After isolating the ground state
contribution of the scalar tetraquark state, we get the following result,
\begin{eqnarray}
\Pi(p)&=&\frac{\lambda_{ Z_c}^2}{M_{Z_c}^2-p^2} +\cdots \, \, ,
\end{eqnarray}
where the   pole residue  $\lambda_{Z_c}$ is defined by $\langle 0|J(0)|Z(p)\rangle = \lambda_{Z_c}$ .

 In the following,  we briefly outline  the operator product expansion for the correlation function $\Pi(p)$ in perturbative QCD.  We contract the $u$, $d$ and $c$ quark fields in the correlation function $\Pi(p)$ with Wick theorem, and obtain the result:
\begin{eqnarray}
 \Pi(p)&=&i\epsilon^{ijk}\epsilon^{imn}\epsilon^{i^{\prime}j^{\prime}k^{\prime}}\epsilon^{i^{\prime}m^{\prime}n^{\prime}}\int d^4x e^{ip \cdot x}   \nonumber\\
&&{\rm Tr}\left[ \gamma_{5}C^{kk^{\prime}}(x)\gamma_{5} CU^{jj^{\prime}T}(x)C\right] {\rm Tr}\left[ \gamma_{5} C^{n^{\prime}n}(-x)\gamma_{5} C D^{m^{\prime}mT}(-x)C\right]   \, ,
\end{eqnarray}
where  the $U_{ij}(x)$, $D_{ij}(x)$ and $C_{ij}(x)$ are the full $u$, $d$ and $c$ quark propagators respectively (the $U_{ij}(x)$ and $D_{ij}(x)$ can be written as $S_{ij}(x)$ for simplicity),
 \begin{eqnarray}
S_{ij}(x)&=& \frac{i\delta_{ij}\!\not\!{x}}{ 2\pi^2x^4}-\frac{\delta_{ij}\langle
\bar{q}q\rangle}{12} -\frac{\delta_{ij}x^2\langle \bar{q}g_s\sigma Gq\rangle}{192} -\frac{ig_sG^{a}_{\alpha\beta}t^a_{ij}(\!\not\!{x}
\sigma^{\alpha\beta}+\sigma^{\alpha\beta} \!\not\!{x})}{32\pi^2x^2} -\frac{i\delta_{ij}x^2\!\not\!{x}g_s^2\langle \bar{q} q\rangle^2}{7776}\nonumber\\
&&  -\frac{\delta_{ij}x^4\langle \bar{q}q \rangle\langle g_s^2 GG\rangle}{27648} -\frac{1}{8}\langle\bar{q}_j\sigma^{\mu\nu}q_i \rangle \sigma_{\mu\nu}-\frac{1}{4}\langle\bar{q}_j\gamma^{\mu}q_i\rangle \gamma_{\mu }+\cdots \, ,
\end{eqnarray}
\begin{eqnarray}
C_{ij}(x)&=&\frac{i}{(2\pi)^4}\int d^4k e^{-ik \cdot x} \left\{
\frac{\delta_{ij}}{\!\not\!{k}-m_c}
-\frac{g_sG^n_{\alpha\beta}t^n_{ij}}{4}\frac{\sigma^{\alpha\beta}(\!\not\!{k}+m_c)+(\!\not\!{k}+m_c)
\sigma^{\alpha\beta}}{(k^2-m_c^2)^2}\right.\nonumber\\
&&\left. +\frac{g_s D_\alpha G^n_{\beta\lambda}t^n_{ij}(f^{\lambda\beta\alpha}+f^{\lambda\alpha\beta}) }{3(k^2-m_c^2)^4}-\frac{g_s^2 (t^at^b)_{ij} G^a_{\alpha\beta}G^b_{\mu\nu}(f^{\alpha\beta\mu\nu}+f^{\alpha\mu\beta\nu}+f^{\alpha\mu\nu\beta}) }{4(k^2-m_c^2)^5}+\cdots\right\} \, ,\nonumber\\
f^{\lambda\alpha\beta}&=&(\!\not\!{k}+m_c)\gamma^\lambda(\!\not\!{k}+m_c)\gamma^\alpha(\!\not\!{k}+m_c)\gamma^\beta(\!\not\!{k}+m_c)\, ,\nonumber\\
f^{\alpha\beta\mu\nu}&=&(\!\not\!{k}+m_c)\gamma^\alpha(\!\not\!{k}+m_c)\gamma^\beta(\!\not\!{k}+m_c)\gamma^\mu(\!\not\!{k}+m_c)\gamma^\nu(\!\not\!{k}+m_c)\, ,
\end{eqnarray}
and  $t^n=\frac{\lambda^n}{2}$, the $\lambda^n$ is the Gell-Mann matrix,  $D_\alpha=\partial_\alpha-ig_sG^n_\alpha t^n$ \cite{Reinders85}, then compute  the integrals both in the coordinate and momentum spaces to obtain the correlation function $\Pi(p)$ therefore the QCD spectral density.
In Eq.(7), we retain the terms $\langle\bar{q}_j\sigma_{\mu\nu}q_i \rangle$ and $\langle\bar{q}_j\gamma_{\mu}q_i\rangle$ originate from the Fierz re-arrangement of the $\langle q_i \bar{q}_j\rangle$ to  absorb the gluons  emitted from the heavy quark lines  so as to extract the mixed condensate and four-quark condensate $\langle\bar{q}g_s\sigma G q\rangle$ and $g_s^2\langle\bar{q}q\rangle^2$, respectively.

 Once the analytical expression  is obtained,  we can take the
quark-hadron duality below the continuum threshold  $s_0$ and perform Borel transform  with respect to
the variable $P^2=-p^2$ to obtain  the following QCD sum rule:
\begin{eqnarray}
\lambda^2_{Z_c}\, \exp\left(-\frac{M^2_{Z_c}}{T^2}\right)= \int_{4m_c^2}^{s_0} ds\, \rho(s) \, \exp\left(-\frac{s}{T^2}\right) \, ,
\end{eqnarray}
where
\begin{eqnarray}
\rho(s)&=&\rho_{0}(s)+\rho_{3}(s) +\rho_{4}(s)+\rho_{5}(s)+\rho_{6}(s)+\rho_{7}(s) +\rho_{8}(s)+\rho_{10}(s)\, ,
\end{eqnarray}

\begin{eqnarray}
\rho_{0}(s)&=&\frac{1}{512\pi^6}\int_{y_i}^{y_f}dy \int_{z_i}^{1-y}dz \, yz\, (1-y-z)^3\left(s-\overline{m}_c^2\right)^2\left(7s^2-6s\overline{m}_c^2+\overline{m}_c^4 \right)    \, ,
\end{eqnarray}

\begin{eqnarray}
\rho_{3}(s)&=&-\frac{m_c\langle \bar{q}q\rangle}{16\pi^4}\int_{y_i}^{y_f}dy \int_{z_i}^{1-y}dz \, (y+z)(1-y-z)\left(s-\overline{m}_c^2\right)\left(2s-\overline{m}_c^2\right)  \, ,
\end{eqnarray}

\begin{eqnarray}
\rho_{4}(s)&=&-\frac{m_c^2}{384\pi^4} \langle\frac{\alpha_s GG}{\pi}\rangle\int_{y_i}^{y_f}dy \int_{z_i}^{1-y}dz \left( \frac{z}{y^2}+\frac{y}{z^2}\right)(1-y-z)^3 \nonumber\\
&&\left\{ 2s-\overline{m}_c^2+\frac{\overline{m}_c^4}{6}\delta\left(s-\overline{m}_c^2\right)\right\} \nonumber\\
&&+\frac{1}{512\pi^4} \langle\frac{\alpha_s GG}{\pi}\rangle\int_{y_i}^{y_f}dy \int_{z_i}^{1-y}dz \left( y+z\right)(1-y-z)^2 \left( 10s^2-12s\overline{m}_c^2+3\overline{m}_c^4\right)  \, ,
\end{eqnarray}

\begin{eqnarray}
\rho_{5}(s)&=&\frac{m_c\langle \bar{q}g_s\sigma Gq\rangle}{64\pi^4}\int_{y_i}^{y_f}dy \int_{z_i}^{1-y}dz  \, (y+z) \left(3s-2\overline{m}_c^2 \right) \nonumber\\
&&-\frac{m_c\langle \bar{q}g_s\sigma Gq\rangle}{64\pi^4}\int_{y_i}^{y_f}dy \int_{z_i}^{1-y}dz  \, \left( \frac{y}{z}+\frac{z}{y}\right) (1-y-z) \left(3s-2\overline{m}_c^2 \right)     \, ,
\end{eqnarray}

\begin{eqnarray}
\rho_{6}(s)&=&\frac{m_c^2\langle\bar{q}q\rangle^2}{12\pi^2}\int_{y_i}^{y_f}dy   +\frac{g_s^2\langle\bar{q}q\rangle^2}{108\pi^4}\int_{y_i}^{y_f}dy \int_{z_i}^{1-y}dz\, yz \left\{2s-\overline{m}_c^2 +\frac{\overline{m}_c^4}{6}\delta\left(s-\overline{m}_c^2 \right)\right\}\nonumber\\
&&-\frac{g_s^2\langle\bar{q}q\rangle^2}{512\pi^4}\int_{y_i}^{y_f}dy \int_{z_i}^{1-y}dz \, (1-y-z)\left\{ 2\left(\frac{z}{y}+\frac{y}{z} \right)\left(3s-2\overline{m}_c^2 \right)+\left(\frac{z}{y^2}+\frac{y}{z^2} \right)\right.\nonumber\\
&&\left.m_c^2\left[ 2+ \overline{m}_c^2\delta\left(s-\overline{m}_c^2 \right)\right] \right\} \nonumber\\
&&-\frac{g_s^2\langle\bar{q}q\rangle^2}{3888\pi^4}\int_{y_i}^{y_f}dy \int_{z_i}^{1-y}dz \, (1-y-z)\left\{  3\left(\frac{z}{y}+\frac{y}{z} \right)\left(3s-2\overline{m}_c^2 \right)+\left(\frac{z}{y^2}+\frac{y}{z^2} \right)\right. \nonumber\\
&&\left.m_c^2\left[ 2+\overline{m}_c^2\delta\left(s-\overline{m}_c^2\right)\right]+(y+z)\left[12\left(2s-\overline{m}_c^2\right) +2\overline{m}_c^4\delta\left(s-\overline{m}_c^2\right)\right] \right\}\, ,
\end{eqnarray}

\begin{eqnarray}
\rho_7(s)&=&\frac{m_c^3\langle\bar{q}q\rangle}{288\pi^2  }\langle\frac{\alpha_sGG}{\pi}\rangle\int_{y_i}^{y_f}dy \int_{z_i}^{1-y}dz \left(\frac{y}{z^3}+\frac{z}{y^3}+\frac{1}{y^2}+\frac{1}{z^2}\right)(1-y-z)\nonumber\\
&&\left(1+\frac{ \overline{m}_c^2}{T^2}\right) \delta\left(s-\overline{m}_c^2\right)\nonumber\\
&&-\frac{m_c\langle\bar{q}q\rangle}{96\pi^2}\langle\frac{\alpha_sGG}{\pi}\rangle\int_{y_i}^{y_f}dy \int_{z_i}^{1-y}dz \left(\frac{y}{z^2}+\frac{z}{y^2}\right)(1-y-z)  \left\{2+\overline{m}_c^2\delta\left(s-\overline{m}_c^2\right) \right\}\nonumber\\
&&-\frac{m_c\langle\bar{q}q\rangle}{96\pi^2}\langle\frac{\alpha_sGG}{\pi}\rangle\int_{y_i}^{y_f}dy \int_{z_i}^{1-y}dz\left\{2+ \overline{m}_c^2 \delta\left(s-\overline{m}_c^2\right) \right\} \nonumber\\
&&-\frac{m_c\langle\bar{q}q\rangle}{576\pi^2}\langle\frac{\alpha_sGG}{\pi}\rangle\int_{y_i}^{y_f}dy \left\{2+ \widetilde{m}_c^2 \, \delta \left(s-\widetilde{m}_c^2\right) \right\}\, ,
\end{eqnarray}

\begin{eqnarray}
\rho_8(s)&=&-\frac{m_c^2\langle\bar{q}q\rangle\langle\bar{q}g_s\sigma Gq\rangle}{24\pi^2}\int_0^1 dy \left(1+\frac{\widetilde{m}_c^2}{T^2} \right)\delta\left(s-\widetilde{m}_c^2\right)\nonumber \\
&&+\frac{ m_c^2\langle\bar{q}q\rangle\langle\bar{q}g_s\sigma Gq\rangle}{48\pi^2}\int_{0}^{1} dy \frac{1}{y(1-y)}\delta\left(s-\widetilde{m}_c^2\right)
 \, ,
\end{eqnarray}

\begin{eqnarray}
\rho_{10}(s)&=&\frac{m_c^2\langle\bar{q}g_s\sigma Gq\rangle^2}{192\pi^2T^6}\int_0^1 dy \, \widetilde{m}_c^4 \, \delta \left( s-\widetilde{m}_c^2\right)
\nonumber \\
&&-\frac{m_c^4\langle\bar{q}q\rangle^2}{216T^4}\langle\frac{\alpha_sGG}{\pi}\rangle\int_0^1 dy  \left\{ \frac{1}{y^3}+\frac{1}{(1-y)^3}\right\} \delta\left( s-\widetilde{m}_c^2\right)\nonumber\\
&&+\frac{m_c^2\langle\bar{q}q\rangle^2}{72T^2}\langle\frac{\alpha_sGG}{\pi}\rangle\int_0^1 dy  \left\{ \frac{1}{y^2}+\frac{1}{(1-y)^2}\right\} \delta\left( s-\widetilde{m}_c^2\right)\nonumber\\
&&-\frac{m_c^2\langle\bar{q}g_s\sigma Gq\rangle^2}{192 \pi^2T^4} \int_0^1 dy   \frac{1}{y(1-y)}  \widetilde{m}_c^2 \, \delta\left( s-\widetilde{m}_c^2\right)\nonumber\\
&&+\frac{m_c^2\langle\bar{q}g_s\sigma Gq\rangle^2}{384 \pi^2T^2} \int_0^1 dy   \frac{1}{y(1-y)}   \delta\left( s-\widetilde{m}_c^2\right)\nonumber \\
&&+\frac{m_c^2\langle\bar{q} q\rangle^2}{216 T^6}\langle\frac{\alpha_sGG}{\pi}\rangle\int_0^1 dy \, \widetilde{m}_c^4 \, \delta \left( s-\widetilde{m}_c^2\right) \, ,
\end{eqnarray}
the subscripts  $0$, $3$, $4$, $5$, $6$, $7$, $8$, $10$ denote the dimensions of the  vacuum condensates, $y_{f}=\frac{1+\sqrt{1-4m_c^2/s}}{2}$,
$y_{i}=\frac{1-\sqrt{1-4m_c^2/s}}{2}$, $z_{i}=\frac{y
m_c^2}{y s -m_c^2}$, $\overline{m}_c^2=\frac{(y+z)m_c^2}{yz}$,
$ \widetilde{m}_c^2=\frac{m_c^2}{y(1-y)}$, $\int_{y_i}^{y_f}dy \to \int_{0}^{1}dy$, $\int_{z_i}^{1-y}dz \to \int_{0}^{1-y}dz$ when the $\delta$ functions $\delta\left(s-\overline{m}_c^2\right)$ and $\delta\left(s-\widetilde{m}_c^2\right)$ appear.  We take into account the vacuum condensates which are
vacuum expections of the operators  of the orders $\mathcal{O}( \alpha_s^{k})$ with $k\leq 1$ consistently.

 Differentiate   Eq.(9) with respect to  $\frac{1}{T^2}$, then eliminate the
 pole residues $\lambda_{Z_c}$, we obtain the QCD sum rule for
 the mass of the scalar    tetraquark state,
 \begin{eqnarray}
 M^2_{Z_c}= \frac{\int_{4m_c^2}^{s_0} ds\frac{d}{d \left(-1/T^2\right)}\rho(s)\exp\left(-\frac{s}{T^2}\right)}{\int_{4m_c^2}^{s_0} ds \rho(s)\exp\left(-\frac{s}{T^2}\right)}\, .
\end{eqnarray}

In the following, we  perform Fierz re-arrangement to the current $J$ both in the color and Dirac-spinor  spaces to obtain the  result,
\begin{eqnarray}
J&=&\frac{1}{4}\left\{\,-\bar{c} c\,\bar{d} u+\bar{c}i\gamma_5 c\,\bar{d}i\gamma_5 u-\bar{c} \gamma^\mu c\,\bar{d}\gamma_\mu u-\bar{c} \gamma^\mu\gamma_5 c\,\bar{d}\gamma_\mu\gamma_5 u+\frac{1}{2}\bar{c}\sigma_{\mu\nu} c\,\bar{d}\sigma^{\mu\nu} u\right. \nonumber\\
&&\left.+\bar{c} u\,\bar{d} c-\bar{c}i\gamma_5 u\,\bar{d}i\gamma_5 c+\bar{c} \gamma^\mu u\,\bar{d}\gamma_\mu c+\bar{c} \gamma^\mu\gamma_5 u\,\bar{d}\gamma_\mu\gamma_5 c-\frac{1}{2}\bar{c}\sigma_{\mu\nu} u\,\bar{d}\sigma^{\mu\nu} c  \,\right\} \, ,
\end{eqnarray}
the components  couple  to the meson pairs  $\chi_{c0}a_0^{+}(980)$, $\eta_c\pi^{+}$, $J/\psi \rho^{+}$, $\chi_{c1}\pi^+$, $\chi_{c1}a_1^+(1260)$, $h_c h_1^+(1170)$, $(D_0(2400)\bar{D}_0(2400))^+$,
$(D\bar{D})^+$, $(D^*\bar{D}^*)^+$, $(D_1(2420)\bar{D}_1(2420))^+$, $(D_1(2430)\bar{D}_1(2430))^+$, respectively. The strong decays
\begin{eqnarray}
Z_c^{\pm}(0^{++}) &\to& \chi_{c0}a_0^{\pm}(980)\, , \, \eta_c\pi^{\pm}\, , \, J/\psi \rho^{\pm}\, , \, \chi_{c1}\pi^\pm\, , \, \chi_{c1}a_1^\pm(1260)\, , \, h_c h_1^\pm(1170)\, , \,(D_0(2400)\bar{D}_0(2400))^\pm \, , \, \nonumber\\
&& (D\bar{D})^\pm \, , \, (D^*\bar{D}^*)^\pm \, , \, (D_1(2420)\bar{D}_1(2420))^\pm \, , \, (D_1(2430)\bar{D}_1(2430))^\pm \, ,
\end{eqnarray}
are Okubo-Zweig-Iizuka  super-allowed, if they are kinematically allowed.
 The
diquark-antidiquark type tetraquark state can be taken as a special superposition of a series of  meson-meson pairs, and embodies  the net effects. The decays to its components (meson-meson pairs) are Okubo-Zweig-Iizuka super-allowed, but the re-arrangements in the color-space are non-trivial \cite{Nielsen3900-1,Nielsen3900-2}.

The numerical analysis indicates that the ground state mass of the $S\bar{S}$-type scalar tetraquark state is about $3.82\,\rm{GeV}$,
the strong decays
\begin{eqnarray}
Z_c^{\pm}(0^{++}) &\to&  \eta_c\pi^{\pm}\, , \, \chi_{c1}\pi^\pm\, , \,  (D\bar{D})^\pm  \, ,
\end{eqnarray}
are  kinematically  allowed. The decay $Z_c^{\pm}(0^{++}) \to \chi_{c1}\pi^\pm$ takes place through relative P-wave and is kinematically suppressed.

Now we write down the  three-point correlation functions
$\Pi_{1}(p,q)$ and $\Pi_{2}(p,q)$ to study the strong decays $Z_c^{\pm}(0^{++}) \to  \eta_c\pi^{\pm}\, , \,   (D\bar{D})^\pm  $,
\begin{eqnarray}
\Pi_{1}(p,q)&=&i^2\int d^4xd^4y e^{ip \cdot x}e^{iq \cdot y}\langle 0|T\left\{J_{\eta_c}(x)J_{\pi}(y)J(0)\right\}|0\rangle\, , \nonumber \\
\Pi_{2}(p,q)&=&i^2\int d^4xd^4y e^{ip\cdot x}e^{iq\cdot y}\langle 0|T\left\{J_{D^-}(x)J_{D^0}(y)J(0)\right\}|0\rangle \, ,
\end{eqnarray}
where the currents
\begin{eqnarray}
J_{\eta_c}(x)&=&\bar{c}(x)i\gamma_5 c(x) \, ,\nonumber \\
J_\pi(y)&=&\bar{u}(y)i\gamma_5 d(y) \, ,\nonumber \\
J_{D^-}(x)&=&\bar{c}(x)i\gamma_5 d(x) \, ,\nonumber \\
J_{D^0}(y)&=&\bar{u}(y)i\gamma_5 c(y) \, ,
\end{eqnarray}
interpolate the mesons $\eta_c$, $\pi$, $D^-$, $D^0$, respectively.

We insert  a complete set of intermediate hadronic states with
the same quantum numbers as the current operators into the three-point
correlation functions $\Pi_{1}(p,q)$ and $\Pi_{2}(p,q)$ and  isolate the ground state
contributions to obtain the following results,
\begin{eqnarray}
\Pi_{1}(p,q)&=& \frac{f_{\pi}M_{\pi}^2f_{\eta_c}M_{\eta_c}^2\lambda_{Z_c}G_{Z_c\eta_c \pi}}{2(m_u+m_d)m_c} \frac{-q\cdot p}{(M_{Z_c}^2-p^{\prime2})(M_{\eta_c}^2-p^2)(M_{\pi}^2-q^2)} +\cdots \, , \nonumber\\
\Pi_{2}(p,q)&=& \frac{f_{D}^2M_{D}^4\lambda_{Z_c}G_{Z_cDD}}{(m_c+m_q)^2} \frac{-q\cdot p}{(M_{Z_c}^2-p^{\prime2})(M_{D}^2-p^2)(M_{D}^2-q^2)} +\cdots \, ,
\end{eqnarray}
where $p^\prime=p+q$, the $f_D$, $f_{\eta_c}$ and $f_{\pi}$ are the decay constants of the mesons  $D$, $\eta_c$ and $\pi$, respectively, the $G_{Z_c\eta_c\pi}$ and $G_{Z_cDD}$ are the hadronic coupling constants. In the following, we write down the definitions,
\begin{eqnarray}
\langle0|J_{\eta_c}(0)|\eta_c(p)\rangle&=&\frac{f_{\eta_c}M_{\eta_c}^2}{2m_c} \,\, , \nonumber \\
\langle0|J_{\pi}(0)|\pi(q)\rangle&=&\frac{f_{\pi}M_{\pi}^2}{m_u+m_d} \,\, ,\nonumber \\
\langle0|J_{D}(0)|D(p/q)\rangle&=&\frac{f_{D}M_{D}^2}{m_c+m_q} \,\, , \\
\langle\eta_c(p)\pi(q)|Z_c(p^{\prime})\rangle&=&-iq\cdot p G_{Z_c\eta_c\pi}(q^2) \, , \nonumber\\
\langle D(p)D(q)|Z_c(p^{\prime})\rangle&=&-iq\cdot p G_{Z_cDD}(q^2) \, .
\end{eqnarray}

\begin{figure}
 \centering
 \includegraphics[totalheight=3cm,width=7cm]{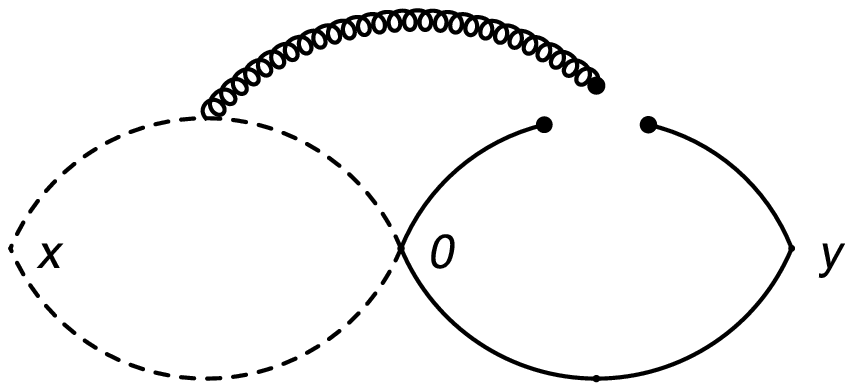}
    \caption{The   connected Feynman diagram contributes to the correlation function $\Pi_{1}(p,q)$, where the dashed and solid lines denote the heavy quark and light quark lines, respectively. Other diagrams obtained  by interchanging of the heavy quark lines or light quark lines are
implied.  }
\end{figure}

\begin{figure}
 \centering
 \includegraphics[totalheight=3cm,width=7cm]{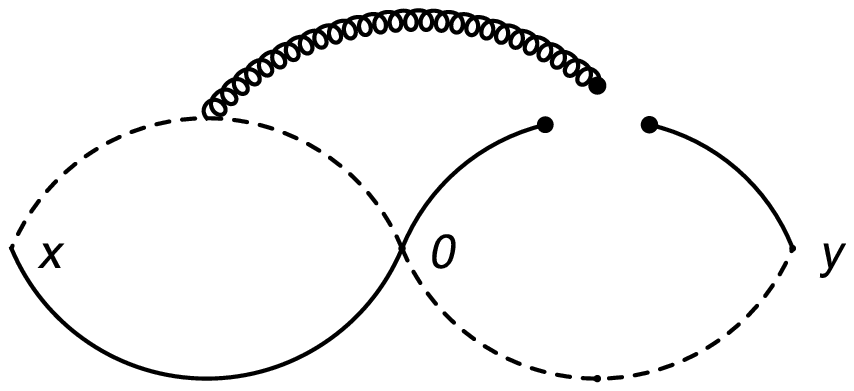}
    \caption{The   connected Feynman diagram contributes to the correlation function $\Pi_{2}(p,q)$, where the dashed and solid lines denote the heavy quark and light quark lines, respectively. Other diagrams obtained  by interchanging of the heavy quark lines and (or) light quark lines are
implied. }
\end{figure}
We carry out the operator product expansion and take into account the color connected Feynman diagrams \cite{Nielsen3900-1,Nielsen3900-2}, and obtain the following results,
\begin{eqnarray}
\Pi_{1}(p,q)&=&-\frac{ m_c\langle \bar{q}g_s\sigma Gq\rangle }{32\pi^2q^2}\int_0^1 dx \frac{q\cdot p}{m_c^2-x(1-x)p^2}+\cdots \, ,
\end{eqnarray}
\begin{eqnarray}
\Pi_{2}(p,q)&=&-\frac{ m_c\langle \bar{q}g_s\sigma Gq\rangle }{64\pi^2}\frac{q \cdot p}{q^2-m_c^2}\int_0^1 dx \frac{1+x}{m_c^2-(1-x)p^2} \nonumber\\
&&-\frac{ m_c\langle \bar{q}g_s\sigma Gq\rangle }{64\pi^2}\frac{q \cdot p}{m_c^2-p^2} \int_0^1 dx  \frac{2-x}{xq^2-m_c^2}+\cdots \, .
\end{eqnarray}
In Fig.1 and Fig.2, we draw the connected Feynman diagrams contribute to the correlation functions $\Pi_{1}(p,q)$ and $\Pi_{2}(p,q)$, respectively.
The $\Pi_{1}(p,q)$ and $\Pi_{2}(p,q)$ can be expanded in terms of the $\cos\theta$, $\Pi_{1/2}(p,q)=\Pi^0(p^2,q^2)+\Pi^1(p^2,q^2)\cos\theta+\Pi^2(p^2,q^2)\cos^2\theta+\cdots$,  at the QCD side, where the $\theta$ is the included angle of the Euclidean momenta $p$ and $q$, i.e. $\cos\theta=p\cdot{q}/\sqrt{q^2p^2}$.
There exists only one term ($\Pi^1(p^2,q^2)\cos\theta$) for the $\Pi_{1}(p,q)$, while there exist two terms ($\Pi^0(p^2,q^2)$ and $\Pi^1(p^2,q^2)\cos\theta$) for
the $\Pi_{2}(p,q)$. At the phenomenological side, the hadronic coupling constants $G_{SPP^\prime}(p,q)$ have the possible forms  $G_{SPP^\prime}^0$,  $G_{SPP^\prime}^1\cos\theta$, $G_{SPP^\prime}^2\cos^2\theta$, $\cdots$, where the $S$ denotes the scalar mesons, the $P$ and $P^{\prime}$ denote the pseudoscalar mesons.   In the present case, it is better to choose the form $G_{SPP^\prime}^1\cos\theta$, as  the correlation functions $\Pi_{1}(p,q)$ and
$\Pi_{2}(p,q)$ both have the term proportional to $\cos\theta$ at the QCD side.
The $\cos\theta$ is the pertinent tensor structure, as the correlation functions $\Pi_{1}(p,q)$ and $\Pi_{2}(p,q)$ should have the same tensor structure at the  phenomenological side.
There exists some shortcoming,
 if we choose the form $G_{SPP^\prime}^0$ and take the replacement  $2p\cdot{q}=p^{\prime2}-p^2-q^2$, then set $p^2=p^{\prime 2}$ and perform  the Borel transform with respect to the variable   $P^2=-p^2$, as the $p$, $q$ and $p^{\prime}$ are not independent variables, the $\cos\theta$ cannot be replaced.

Once the analytical expressions of the correlation functions $\Pi_{1}(p,q)$ and $\Pi_{2}(p,q)$ at both the QCD side and hadron  side are obtained, we perform  the Borel transform with respect to the variable   $P^2=-p^2$ by setting $p^2=p^{\prime 2}$, then take the quark-hadron duality   and obtain the following QCD sum rules,
\begin{eqnarray}
&&\frac{f_{\pi}M_{\pi}^2f_{\eta_c}M_{\eta_c}^2\lambda_{Z_c}G_{Z_c\eta_c \pi}}{2(m_u+m_d)m_c(M_{Z_c}^2-M_{\eta_c}^2)} \left\{ \exp\left(-\frac{M_{\eta_c}^2}{T^2} \right)-\exp\left(-\frac{M_{Z_c}^2}{T^2} \right)\right\}+C_{Z_c\eta_c \pi} \exp\left(-\frac{s_0}{T^2} \right) \nonumber\\
&&=-\frac{m_c\langle\bar{q}g_s\sigma Gq\rangle}{32\pi^2}\frac{Q^2+M_{\pi}^2}{Q^2} \int_0^1 dx \frac{ 1}{x(1-x)}\exp\left( -\frac{m_c^2}{x(1-x)T^2}\right)\, ,
\end{eqnarray}
\begin{eqnarray}
&&\frac{f_{D}^2M_{D}^4\lambda_{Z_c}G_{Z_cDD}}{(m_c+m_q)^2 (M_{Z_c}^2-M_{D}^2)} \left\{ \exp\left(-\frac{M_{D}^2}{T^2} \right)-\exp\left(-\frac{M_{Z_c}^2}{T^2} \right)\right\}+C_{Z_cDD} \exp\left(-\frac{s_0}{T^2} \right) \nonumber\\
&&=-\frac{m_c\langle\bar{q}g_s\sigma Gq\rangle}{64\pi^2}(Q^2+M_{D}^2)\int_0^1 dx \left\{\frac{1}{Q^2+m_c^2}\frac{1+x}{(1-x)}\exp\left( -\frac{m_c^2}{(1-x)T^2}\right) \right. \nonumber\\
&&\left. +\frac{2-x}{xQ^2+m_c^2}\exp\left( -\frac{m_c^2}{T^2}\right) \right\}\, ,
\end{eqnarray}
where the $s_0$ is the continuum threshold parameter for the $Z_c$, and the $C_{Z_c\eta_c\pi}$ and $C_{Z_cDD}$ are unknown  parameters introduced to take into account
the single-pole contributions associated with pole-continuum
transitions. In numerical analysis, we will denote the right sides of Eqs.(30-31) as $F_1(Q^2)$ and $F_2(Q^2)$ respectively.   In the three-point QCD sum rules, the single-pole contributions  are not suppressed if a single
Borel transform is taken.

\section{Numerical results and discussions}
The vacuum condensates are taken to be the standard values
$\langle\bar{q}q \rangle=-(0.24\pm 0.01\, \rm{GeV})^3$,
$\langle\bar{q}g_s\sigma G q \rangle=m_0^2\langle \bar{q}q \rangle$,
$m_0^2=(0.8 \pm 0.1)\,\rm{GeV}^2$, $\langle \frac{\alpha_s
GG}{\pi}\rangle=(0.33\,\rm{GeV})^4 $    at the energy scale  $\mu=1\, \rm{GeV}$
\cite{SVZ79-1,SVZ79-2,Reinders85,Ioffe2005-1,Ioffe2005-2}.
The quark condensate and mixed quark condensate evolve with the   renormalization group equation,
$\langle\bar{q}q \rangle(\mu)=\langle\bar{q}q \rangle(Q)\left[\frac{\alpha_{s}(Q)}{\alpha_{s}(\mu)}\right]^{\frac{4}{9}}$ and
 $\langle\bar{q}g_s \sigma Gq \rangle(\mu)=\langle\bar{q}g_s \sigma Gq \rangle(Q)\left[\frac{\alpha_{s}(Q)}{\alpha_{s}(\mu)}\right]^{\frac{2}{27}}$.

The hadronic input parameters are taken as $M_{\pi}=0.13957\,\rm{GeV}$, $f_{\pi}=0.130\,\rm{GeV}$,
$M_{D^\pm}=1.8695\,\rm{GeV}$, $M_{D^0}=1.86491\,\rm{GeV}$, $f_{D}=0.208\,\rm{GeV}$,  $M_{\eta_c}=2.9837\,\rm{GeV}$,
$f_{\eta_c}=0.350 \,\rm{GeV}$ \cite{PDG,WangJHEP,VANovikov}.

We take the values $m_u({\mu=\rm 1GeV})=m_d({\mu=\rm 1GeV})=m_q({\mu=\rm 1GeV})=0.006\,\rm{GeV}$ from the Gell-Mann-Oakes-Renner relation, and choose the $\overline{MS}$ mass $m_{c}(m_c)=(1.275\pm0.025)\,\rm{GeV}$
 from the Particle Data Group \cite{PDG}, and take into account
the energy-scale dependence of  the $\overline{MS}$ masses  from the renormalization group equation,
\begin{eqnarray}
m_q(\mu)&=&m_q({\rm1GeV})\left[\frac{\alpha_{s}(\mu)}{\alpha_{s}({\rm1GeV})}\right]^{\frac{4}{9}} \, ,\nonumber\\
m_c(\mu)&=&m_c(m_c)\left[\frac{\alpha_{s}(\mu)}{\alpha_{s}(m_c)}\right]^{\frac{12}{25}} \, ,\nonumber\\
\alpha_s(\mu)&=&\frac{1}{b_0t}\left[1-\frac{b_1}{b_0^2}\frac{\log t}{t} +\frac{b_1^2(\log^2{t}-\log{t}-1)+b_0b_2}{b_0^4t^2}\right]\, ,
\end{eqnarray}
  where $t=\log \frac{\mu^2}{\Lambda^2}$, $b_0=\frac{33-2n_f}{12\pi}$, $b_1=\frac{153-19n_f}{24\pi^2}$, $b_2=\frac{2857-\frac{5033}{9}n_f+\frac{325}{27}n_f^2}{128\pi^3}$,  $\Lambda=213\,\rm{MeV}$, $296\,\rm{MeV}$  and  $339\,\rm{MeV}$ for the flavors  $n_f=5$, $4$ and $3$, respectively  \cite{PDG}.

Now we study the mass and pole residue of the $S\bar{S}$ type scalar tetraquark state.
  We impose
the two criteria (pole dominance and convergence of the operator product
expansion) on the hidden charmed tetraquark state to choose the Borel
parameter $T^2$ and threshold parameter $s_0$.

In the heavy quark limit, the $c$ (and $b$) quark can be taken as a static well potential,
which binds the light quark $q^{\prime}$ to form a diquark in the color antitriplet channel or binds the light antiquark $\bar{q}$ to form a meson in the color singlet channel (or a meson-like state in the color octet  channel). Then the heavy tetraquark states  are characterized by the effective heavy quark masses ${\mathbb{M}}_Q$ (or constituent quark masses) and the virtuality $V=\sqrt{M^2_{X/Y/Z}-(2{\mathbb{M}}_Q)^2}$ (or bound energy not as robust). It is natural to take the energy  scale $\mu=V$,
 the formula works well for the  $X(3872)$, $Z_c(3885)$,
$Z_c(3900)$,  $Z_c(4020)$, $Z_c(4025)$, $Z(4050)$, $Z(4250)$, $Y(4360)$, $Z(4430)$, $Y(4630)$, $Y(4660)$, $Z_b(10610)$  and $Z_b(10650)$ in the  scenario of  tetraquark  states \cite{WangHuangTao-1,WangHuangTao-2,WangHuangTao-3,Wang-Cu-Cu}.
The relation
\begin{eqnarray}
M^2_{X/Y/Z}=(2{\mathbb{M}}_c)^2+\mu^2 \, ,
\end{eqnarray}
with the value ${\mathbb{M}}_c=1.8\,\rm{GeV}$ determined in previous works \cite{WangHuangTao-1,WangHuangTao-2,WangHuangTao-3,Wang-Cu-Cu} puts  a strong constraint on the masses of the possible tetraquark states.

The mass gaps between the ground states  and the first radial excited states are usually taken as $(0.4-0.6)\,\rm{GeV}$, for example,
the  $Z(4430)$ is tentatively assigned as the first radial excitation of the $Z_c(3900)$ according to the
analogous decays,
\begin{eqnarray}
Z_c(3900)^\pm&\to&J/\psi\pi^\pm\, , \nonumber \\
Z(4430)^\pm&\to&\psi^\prime\pi^\pm\, ,
\end{eqnarray}
and the mass differences $M_{Z(4430)}-M_{Z_c(3900)}=576\,\rm{MeV}$, $M_{\psi^\prime}-M_{J/\psi}=589\,\rm{MeV}$ \cite{Maiani-2014,Nielsen-1401,Wang4430}.
The relation
\begin{eqnarray}
\sqrt{s_0}&=&M_{X/Y/Z}+(0.4-0.6)\,\rm{GeV} \, ,
\end{eqnarray}
puts another strong constraint on the masses of the possible tetraquark states.

 In calculations, we observe that
\begin{eqnarray}
\mu\uparrow  \, \, \, \, \,  M_Z \downarrow \, ,\nonumber\\
\mu\downarrow  \, \, \, \, \,  M_Z \uparrow \, ,
\end{eqnarray}
from the QCD sum rule in Eq.(19). While Eq.(33) indicates that
\begin{eqnarray}
\mu\uparrow  \, \, \, \, \,  M_Z \uparrow \, ,\nonumber\\
\mu\downarrow  \, \, \, \, \,  M_Z \downarrow \, .
\end{eqnarray}
There must be a compromise, which leads  to the optimal energy scale $\mu$, mass $M_Z$ and threshold parameter $s_0$.

In Fig.3,  the contribution of the pole term is plotted with
variations of the threshold parameter $s_0$ and Borel parameter $T^2$ at the energy scale $\mu=1.3\,\rm{GeV}$.
From the figure, we can see that  the value  $\sqrt{s_0}\leq 4.1 \, \rm{GeV}$ is too small to satisfy the pole dominance condition and result in reasonable Borel window.

\begin{figure}
\centering
\includegraphics[totalheight=8cm,width=10cm]{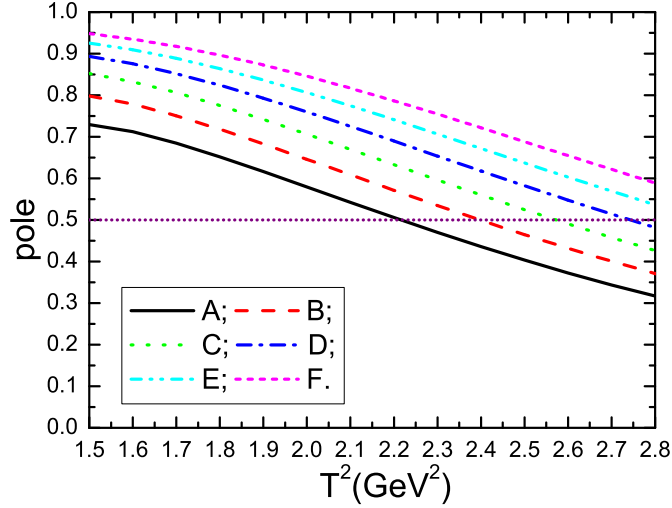}
  \caption{ The pole contribution with variations of the  Borel parameter $T^2$ and threshold parameter $s_0$, where the $A$, $B$, $C$, $D$, $E$, $F$ denote the threshold parameters $\sqrt{s_0}=4.0$,  $4.1$, $4.2$, $4.3$, $4.4$, $4.5\,\rm{GeV}$, respectively. }
\end{figure}

In Fig.4,  the contributions of different terms in the
operator product expansion are plotted with variations of the Borel parameter   $T^2$ for the threshold parameter $\sqrt{s_0}= 4.3 \, \rm{GeV}$ at the energy scale $\mu=1.3\,\rm{GeV}$.
From the figure, we can see that  the $D_0$, $D_3$, $D_5$, $D_6$ and $D_8$, where the $D_i$ denote the contributions of the vacuum condensates of dimensions $D=i$,
play an important  role, while the $D_4$, $D_7$ and $D_{10}$ play a minor important role. At the value $T^2\leq 2.0\,\rm{GeV}$, the $D_3$, $D_5$, $D_6$ and $D_8$ decrease monotonously and quickly with increase of the $T^2$, which cannot lead to stable QCD sum rules. At the value $T^2=(2.2-2.6)\,\rm{GeV}^2$, $D_3\gg |D_5|\gg D_6\gg |D_8|$ and $D_{10}\ll 1\%$, the operator product expansion is well convergent, although $D_0\approx20\%$.
We approximate the continuum spectral density by
$\rho_{QCD}(s)\Theta(s-s_0)$; the contributions of the quark condensate $\langle\bar{q}q\rangle$ and mixed condensate  $\langle\bar{q}g_s\sigma Gq\rangle$   can be very large.

\begin{figure}
\centering
\includegraphics[totalheight=8cm,width=10cm]{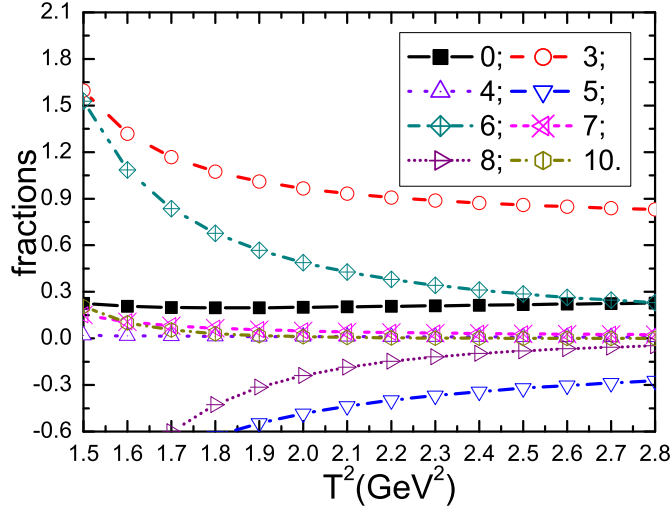}
  \caption{ The contributions of different terms in the operator product expansion  with variations of the  Borel parameter $T^2$, where the 0, 3, 4, 5, 6, 7, 8, 10 denote the dimensions of the vacuum condensates. }
\end{figure}

In this article, we take the  Borel parameter
 $T^2=(2.2-2.6)\,\rm{GeV}^2$,
 the continuum  threshold parameter
 $\sqrt{s_0}=(4.2-4.4)\,\rm{GeV}$ and the energy scale $\mu=1.3\,\rm{GeV}$, the pole dominance is well satisfied.
The Borel parameter, continuum threshold parameter and the pole contribution are shown explicitly in Table 1. The two criteria (pole dominance and convergence of the operator product expansion) of the QCD sum rules are fully satisfied, furthermore, the relations in Eq.(33) and Eq.(35) are also satisfied.

\begin{table}
\begin{center}
\begin{tabular}{|c|c|c|c|c|c|c|c|}\hline\hline
   $J^{PC}$   & $T^2 (\rm{GeV}^2)$ & $\sqrt{s_0} (\rm{GeV})$   & pole         & $M_{Z}(\rm{GeV})$         & $\lambda_{Z}$ \\ \hline
   $0^{++}$   & $2.2-2.6$          & $4.3\pm0.1$               & $(49-74)\%$  & $3.82^{+0.08}_{-0.08}$    & $1.79^{+0.29}_{-0.24}\times10^{-2}\rm{GeV}^5$    \\ \hline
 \hline
\end{tabular}
\end{center}
\caption{ The Borel parameter, continuum threshold parameter, pole contribution, mass  and pole residue of the scalar  tetraquark state. }
\end{table}

\begin{figure}
\centering
\includegraphics[totalheight=8cm,width=10cm]{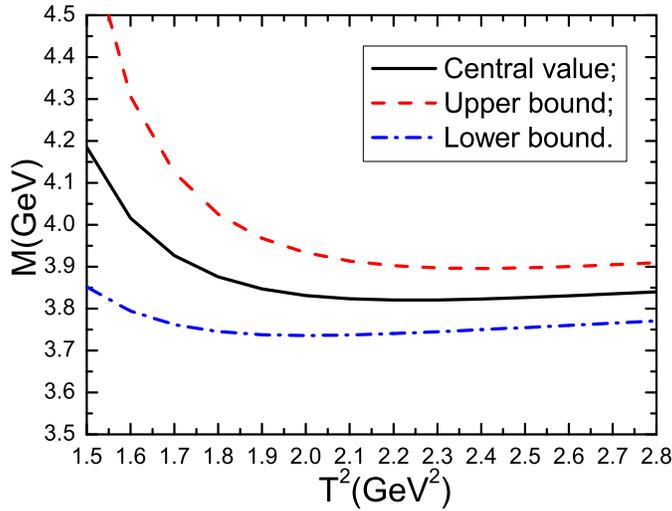}
  \caption{ The mass  with variations of the  Borel parameter $T^2$. }
\end{figure}

Taking into account all uncertainties of the input parameters,
finally we obtain the values of the mass and pole residue of
 the  $S\bar{S}$ type scalar tetraquark state, which are  shown explicitly in Figs.5-6 and Table 1.

The central value of the  present prediction $M_{Z_c}=\left(3.82^{+0.08}_{-0.08}\right)\,\rm{GeV}$ for the $S\bar{S}$ type scalar tetraquark state is smaller than
that of the  $A\bar{A}$ type scalar tetraquark state $M_{J=0}=\left(3.85^{+0.15}_{-0.09}\right)\,\rm{GeV}$ obtained in Ref.\cite{Wang-Cu-Cu}.
The predictions based on the QCD sum rules are consistent with the values $M_{J=0}=3.852\,\rm{GeV}$ and $3.812\,\rm{GeV}$ for the $A\bar{A}$ and $S\bar{S}$ type scalar tetraquark states $cq\bar{c}\bar{q}$ respectively from the quasipotential approach \cite{EFG-2008}.

Now we take the mass $M_{Z_c}$ and pole residue $\lambda_{Z_c}$ as basic input parameters to study the hadronic coupling constants $G_{Z_c\eta_c\pi}$ and $G_{Z_cDD}$, and take the same threshold parameter and Borel parameter as in the QCD sum rule for the mass and pole residue. In calculations, we choose
the unknown parameters  as $C_{Z_c\eta_c\pi}=0.0009\,\rm{GeV}^6 $ and $C_{Z_cDD}=0.0004\,\rm{GeV}^6 $ to obtain stable QCD sum rules with variations of the  Borel parameter $T^2$ at the Borel windows $T^2=(2.2-2.6)\,\rm{GeV}^2$;  the left side and right side of the QCD sum rules coincide.  In fact, it is not necessary to choose the same Borel parameters both in the two-point and three-point QCD sum rules. If we take larger Borel parameter, say $T^2=(2.5-3.0)\,\rm{GeV}^2$ instead of $T^2=(2.2-2.6)\,\rm{GeV}^2$, we should alter the  unknown parameters   $C_{Z_c\eta_c\pi}$ and $C_{Z_cDD} $ slightly, then obtain stable QCD sum rules, the resulting values of the hadronic coupling constants change slightly.

\begin{figure}
\centering
\includegraphics[totalheight=8cm,width=10cm]{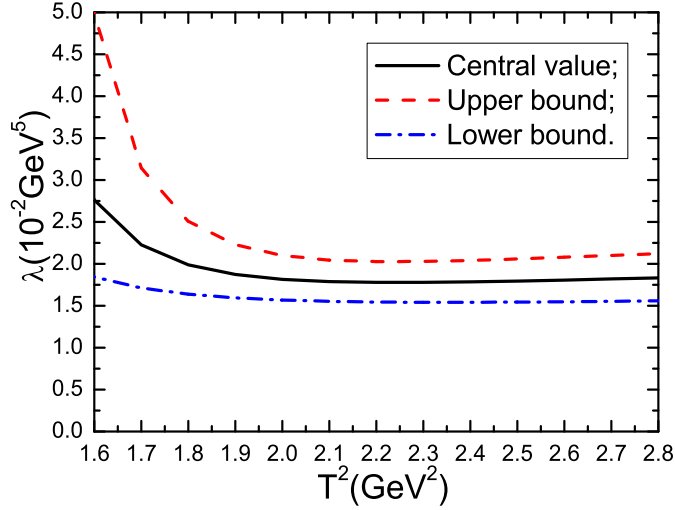}
  \caption{ The pole residue  with variations of the  Borel parameter $T^2$. }
\end{figure}

\begin{figure}
\centering
\includegraphics[totalheight=8cm,width=10cm]{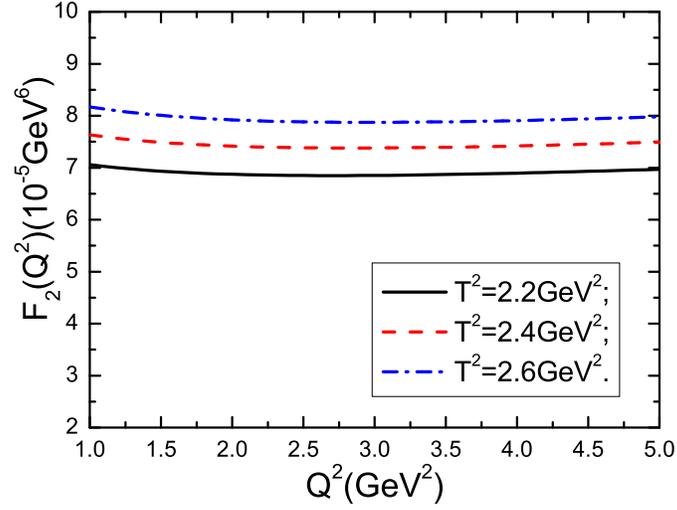}
  \caption{ The central values of the $F_2(Q^2)$  with variations of the $Q^2$. }
\end{figure}

\begin{figure}
\centering
\includegraphics[totalheight=8cm,width=10cm]{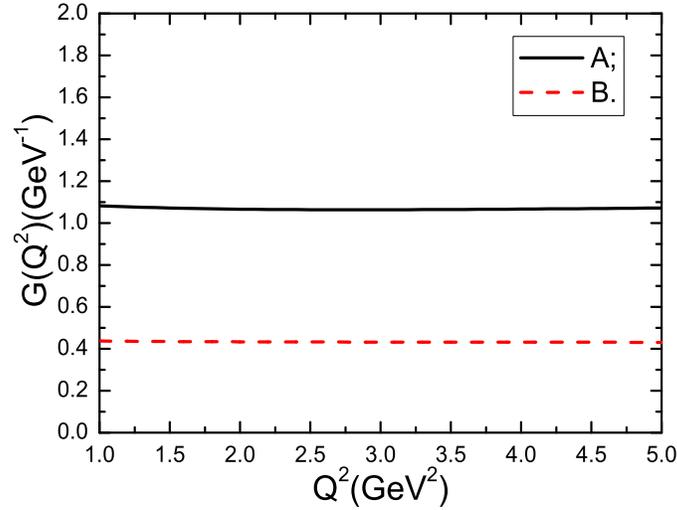}
  \caption{ The central values of the hadronic coupling constants   with variations of the  $Q^2$, where the $A$ and $B$ denote the $G_{Z_cDD}(Q^2)$ and $G_{Z_c\eta_c\pi}(Q^2)$, respectively. }
\end{figure}

Based on Eqs.(30-31), we can study the $Q^2$ dependence of the right side of the QCD sum rules,
\begin{eqnarray}
F_1(Q^2)&\propto&\frac{Q^2+M_{\pi}^2}{Q^2}\approx 1 \, ,
\end{eqnarray}
at the region of large (or intermediate) $Q^2$ due to the tiny mass of the $\pi$, while the $F_2(Q^2)$ has no such simple  $Q^2$ dependence  due to the heavy quark mass $m_c$ and heavy meson mass $M_D$. In the limit $Q^2\to \infty$,
 \begin{eqnarray}
F_2(Q^2)&=&-\frac{m_c\langle\bar{q}g_s\sigma Gq\rangle}{64\pi^2}\int_0^1 dx \left\{\frac{1+x}{1-x}\exp\left( -\frac{m_c^2}{(1-x)T^2}\right)  +\frac{2-x}{x}\exp\left( -\frac{m_c^2}{T^2}\right) \right\}\, ,
\end{eqnarray}
which is independence on $Q^2$.
 In Fig.7, we plot the central values of the  $F_2(Q^2)$ with variations  of the $Q^2$ at the range $Q^2=(1-5)\,\rm{GeV}^2$ for the Borel parameters  $T^2=2.2\,\rm{GeV}^2$, $2.4\,\rm{GeV}^2$ and $2.6\,\rm{GeV}^2$, respectively.
From the figure, we can see that the $Q^2$ dependence of the $F_2(Q^2)$ is rather mild and can be neglected approximately.
The left sides of the QCD sum rules in Eqs.(30-31) have no explicit $Q^2$ dependence, the $Q^2$ dependence is embodied in the right sides of the QCD sum rules ($F_1(Q^2)$ and $F_2(Q^2)$), so the hadronic coupling constants
$G_{Z_c\eta_c\pi}$ and $G_{Z_cDD}$ are  independent on the $Q^2$   in the limit $Q^2\rightarrow \infty$, the conclusion survives even for much smaller $Q^2$, say $Q^2=(1-5)\,\rm{GeV}^2$ according to Eq.(38) and Fig.7.
The central values of the $G_{Z_c\eta_c\pi}(Q^2)$ and $G_{Z_cDD}(Q^2)$ can be fitted to the following constant forms,
\begin{eqnarray}
G_{Z_c\eta_c\pi}(Q^2)&=&0.43\,\rm{GeV}^{-1}\, , \nonumber \\
G_{Z_cDD}(Q^2)&=& 1.06\,\rm{GeV}^{-1} \, ,
\end{eqnarray}
at the region $Q^2=(1-5)\,\rm{GeV}^2$; the uncertainties of the $G_{Z_c\eta_c\pi}$ and $ G_{Z_cDD}$ are about $25\%$ and $18\%$, respectively.
We plot the central values of the hadronic coupling constants $G_{Z_cDD}(Q^2)$ and $G_{Z_c\eta_c\pi}(Q^2)$  with variations of the  $Q^2$ at the region $Q^2=(1-5)\,\rm{GeV}^2$ for the Borel parameter $T^2=2.4\,\rm{GeV}^2$ in Fig.8. From the figure, we can see that the fitted functions  in Eq.(40) are satisfactory.
We extend the coupling constants to the physical regions without difficulty, and calculate the partial decay widths,
\begin{eqnarray}
\Gamma_{Z_c \to \eta_c\pi}&=&\frac{G_{Z_c\eta_c\pi}^2(M_{Z_c}^2-M_{\eta_c}^2-M_{\pi}^2)^2\,p_{\eta_c\pi}}{32\pi M_{Z_c}^2}=(3.0\pm1.5)\,\rm{MeV}\, , \nonumber\\
\Gamma_{Z_c\to DD}&=&\frac{G_{Z_cDD}^2(M_{Z_c}^2-M_{D^+}^2-M_{D^0}^2)^2\,p_{DD}}{32\pi M_{Z_c}^2}=(17.9\pm6.4)\,\rm{MeV}\, ,
\end{eqnarray}
where
\begin{eqnarray}
p_{\eta_c\pi}&=&\frac{\sqrt{\left[M_{Z_c}^2-(M_{\eta_c}+M_{\pi})^2\right]\left[M_{Z_c}^2-(M_{\eta_c}-M_{\pi})^2\right]}}{2M_{Z_c}} \, ,\nonumber\\
p_{DD}&=&\frac{\sqrt{\left[M_{Z_c}^2-(M_{D^+}+M_{D^0})^2\right]\left[M_{Z_c}^2-(M_{D^+}-M_{D^0})^2\right]}}{2M_{Z_c}} \, .
\end{eqnarray}
The total width $\Gamma_{Z_c}$ of the $Z_c(3820)$ can be approximated by $\Gamma_{Z_c \to \eta_c\pi}+\Gamma_{Z_c\to DD}$,  the numerical value is about  $(20.9\pm 6.6)\,\rm{MeV}$. The radiative decay widths can be estimated by  assuming  vector meson dominance, for example,
$\Gamma_{Z_c^\pm \to \gamma\rho^\pm}  \propto\alpha |\Gamma_{Z^\pm_c \to J/\psi^* \rho^\pm}|$ for the radiative decays $Z^\pm_c(3820) \to J/\psi^* \rho^\pm \to \gamma\rho^\pm$,  the partial decay widths  are of the order ${\cal{O}}(\rm {KeV})$ due to the factor $\alpha=\frac{e^2}{4\pi}=\frac{1}{137}$. The strong decays
$Z^\pm_c(3820) \to J/\psi\rho^\pm$ are kinematically forbidden, the values of the $\Gamma_{Z^\pm_c \to J/\psi^* \rho^\pm}$ are complex, so we take $|\Gamma_{Z^\pm_c \to J/\psi^* \rho^\pm}|$.       The  contributions of the radiative decays to the total width $\Gamma_{Z_c}$ are small and can be neglected.

\section{Conclusion}
In this article, we calculate the contributions of the vacuum condensates up to dimension-10  in the operator product expansion,  study the   $S\bar{S}$ type scalar tetraquark state $cq\bar{c}\bar{q}$ in details with the QCD sum rules. In calculations, we search for the optimal Borel parameter and threshold parameter to satisfy    the  energy scale formula $M^2_{Z}=(2{\mathbb{M}}_c)^2+\mu^2$  and the experiential threshold formula $\sqrt{s_0}=M_{Z}+(0.4-0.6)\,\rm{GeV}$, where the $\mu$ is the energy scale of the QCD spectral density, and obtain the values $M_{Z_c}=\left(3.82^{+0.08}_{-0.08}\right)\,\rm{GeV}$ and $\lambda_{Z_c}=\left(1.79^{+0.29}_{-0.24}\right)\times10^{-2}\rm{GeV}^5$. The  central value of the mass of the  $S\bar{S}$ type scalar tetraquark state is smaller than that of the $A\bar{A}$ type scalar tetraquark state, the $S\bar{S}$ type scalar tetraquark state $cq\bar{c}\bar{q}$ maybe the lowest hidden charmed tetraquark state. Furthermore, we calculate  the hadronic coupling constants $G_{Z_c\eta_c\pi}$ and $G_{Z_cDD}$ with the three-point QCD sum rules by taking into account the color-connected diagrams, then study the strong decays  $ Z_c\to \eta_c\pi\, ,\, DD$, and observe that the total width $\Gamma_{Z_c}\approx 21\,\rm{MeV}$.  The present
  predictions can be confronted with the experimental data in the futures at the BESIII, LHCb and Belle-II.

\section*{Acknowledgements}
This  work is supported by National Natural Science Foundation,
Grant Numbers 11375063,  and Natural Science Foundation of Hebei province, Grant Number A2014502017.

\end{document}